\documentclass[proof]{WileyASNA-v1}

\articletype{Article Type}%

\usepackage{ulem}

\definecolor{darkred}{cmyk}{0,1,1,0.45}

\received{XXXX}
\revised{XXXX}
\accepted{XXXX}

\begin{document}

\title{Localizing flares to understand stellar magnetic fields and space weather in exo-systems}

\author[1,2]{Ekaterina Ilin*}

\author[1,2]{Katja Poppenh\"ager}

\author[1]{Juli\'an D. Alvarado-G\'omez}


\authormark{ILIN \textsc{et al}}

\address[1]{\orgname{Leibniz Institute for Astrophysics Potsdam (AIP)}, \orgaddress{An der Sternwarte 16, 14482 Potsdam, \country{Germany}}}
\address[2]{\orgdiv{Institute for Physics and Astronomy}, \orgname{University of Potsdam}, \orgaddress{Karl-Liebknecht-Str. 24/25, 14476 Potsdam \country{Germany}}}


\corres{*\email{eilin@aip.de}}


\abstract{Stars are uniform spheres, but only to first order. The way in which stellar rotation and magnetism break this symmetry places important observational constraints on stellar magnetic fields, and factors in the assessment of the impact of stellar activity on exoplanet atmospheres.

The spatial distribution of flares on the solar surface is well known to be non-uniform, but elusive on other stars. We briefly review the techniques available to recover the loci of stellar flares, and highlight a new method that enables systematic flare localization directly from optical light curves.

We provide an estimate of the number of flares we may be able to localize with the Transiting Exoplanet Survey Satellite (TESS), and show that it is consistent with the results obtained from the first full sky scan of the mission. We suggest that non-uniform flare latitude distributions need to be taken into account in accurate assessments of exoplanet habitability. }



\maketitle


\section{Introduction}\label{sec:intro}

Flares are magnetic explosions that occur in the atmospheres of all stars that possess an outer convection zone. They are high-energy phenomena, during which magnetic re-connection triggers a sudden energy release in form of electromagnetic radiation from radio bands, through optical and UV to soft and hard X-ray emission, each with a characteristic time evolution~\citep{priest2002}. These general features are shared by flares observed on most stars from the Sun, solar analogs down to fully convective dwarfs, and from pre-main sequence to solar age~\citep[see, e.g.,][]{getman2005,walkowicz2011, benz2017,paudel2018}. In our interpretation of stellar flare observations we therefore often extrapolate from our knowledge of the Sun.

On the Sun, flares occur in the vicinity of sunspots in a belt around the solar equator, and almost never above $30^\circ$ latitude~\citep{chen2011}. However, the discovery of polar spots on young sun-like and low mass stars broke the analogy between the solar and stellar situations~\citep{strassmeier2002}. Where flares occur on the surfaces of stars is an open question. 

Locating flaring regions and coronal loops on the stellar sphere has so far been possible only in lucky cases, where flares occurred during spectropolarimetric observations, or eclipses of a binary star. 

The latter occurred in the Algol binary system, where a large X-ray flare on Algol B was occulted by the primary star, and revealed that the emission originated from the south pole of the secondary~\citep{schmitt1999}, which was debated later~\citep{sanz-forcada2007eclipsed}. The event was probably also accompanied by a giant coronal mass ejection~\citep[CME,][]{moschou2017}, and could be associated with a large and stable, pole-oriented coronal loop structure~\citep{peterson2010}.

In another lucky case, \citet{wolter2008} could apply the Doppler Imaging (DI) technique to the Ca II K line to localize a flaring region at $56\pm10^\circ$ latitude in the chromosphere on the K2 star BO Mic. The flare lasted for 4~h, and appeared at the fringes of large spot structures that had also been recovered with DI.

More recently, \citet{berdyugina2017} reconstructed a magnetic field loop that erupted in flare-like bursts on a late M dwarf using spectropolarimetry. Because the inclination of the stellar rotation axis of this star was not well-constrained, they could not pin down what latitudes the loop occupied but recovered its topology regardless.

The serendipity of these discoveries is due to the randomness of flare occurrence times, and the rarity of sufficiently gradual, large flares to which the analysis techniques were sensitive. 

Here, we highlight a new direct method for the systematic localization of flares on the stellar surface. In Section~\ref{sec:howto}, we briefly review previous methods, introduce the generalities of our novel approach, and estimate the number of flares suited for this technique. In Section~\ref{sec:implications}, we discuss how flare loci help us understand the nature of stellar magnetic fields ~(Section~\ref{sec:mf}), and outline the implications for exoplanetary space weather~(Section~\ref{sec:planets}). We conclude with a short summary in Section~\ref{sec:summary}.

\section{How to localize flares on the stellar surface}
\label{sec:howto}

\subsection{Indirectly via Doppler Imaging and Zeeman Doppler Imaging }
Doppler Imaging~\citep[DI,][]{deutsch1958, goncharskii1977} and Zeeman Doppler Imaging~\citep[ZDI,][]{semel1989, donati1989} of stellar magnetic fields have been successfully used to map stellar spot structures and global magnetic field configurations in FGKM dwarfs~\citep{strassmeier2002,morin2008,morin2010,see2019}. 

Although spot size generally increases with flare frequency~\citep{maehara2017}, flares do not correlate directly with rotational phase in stars that show activity induced rotational variability~\citep[i.e., dark spots or bright regions,][]{ramsay2013,doyle2019}. So it is not clear where flares occur on these active stars, although we know that on the Sun, flares are found in the vicinity of spots, and more energetic flares occur in more complex spot groups~\citep{sammis2000}. 

\citet{berdyugina2017} and \citet{wolter2008} used ZDI and DI to more directly recover coronal loops and emission loci, respectively, from flares that occurred during their observation campaigns. However, these methods would require prohibitively expensive long-term spectroscopic monitoring to capture flares on purpose.

\subsection{Indirectly via asteroseismology and light curve inversion}

Activity-driven perturbations in asteroseismic observations can in theory reveal the latitudes of active regions~\citep{gizon2002, papini2015, papini2019}. The method poses that strong magnetic field clusters disturb asteroseismic oscillations that then become visible as characteristic shifts in the power spectrum that trace the location of the clusters.

\citet{berdyugina2005} suggested an alternative approach that combines asteroseismic information on the differential rotation profile of a star with spot-induced variability obtained from time series photometry. It proved more feasible than the asteroseismology-only technique, leading to the first multi-epoch spot latitude survey. The result was a time resolved spot latitude map on a sun-like star, which is known as butterfly diagram in the solar case~\citep{bazot2018}.

If an eclipsing body is present in the system, comparing the stellar light curve of the active star in and out of eclipse at the same rotational phases provides information about the structures covered by the eclipsing body. In systems with low obliquity, a well-defined strip in latitude can be resolved in addition to the longitudes~\citep{huber2009,huber2010}; in sufficiently misaligned systems, even butterfly diagrams can be recovered~\citep{netto2020}. 

Although flare localization is in principle possible with these methods, and is certainly less expensive that Z(DI) because it relies mostly on photometric observations, all three methods are better suited to recover more stable magnetically active structures like starspots, which provide only indirect information about flare location.

\subsection{Directly from optical light curves}
\label{sec:systematic}

\begin{figure}
    \centering
    \includegraphics[width=\hsize]{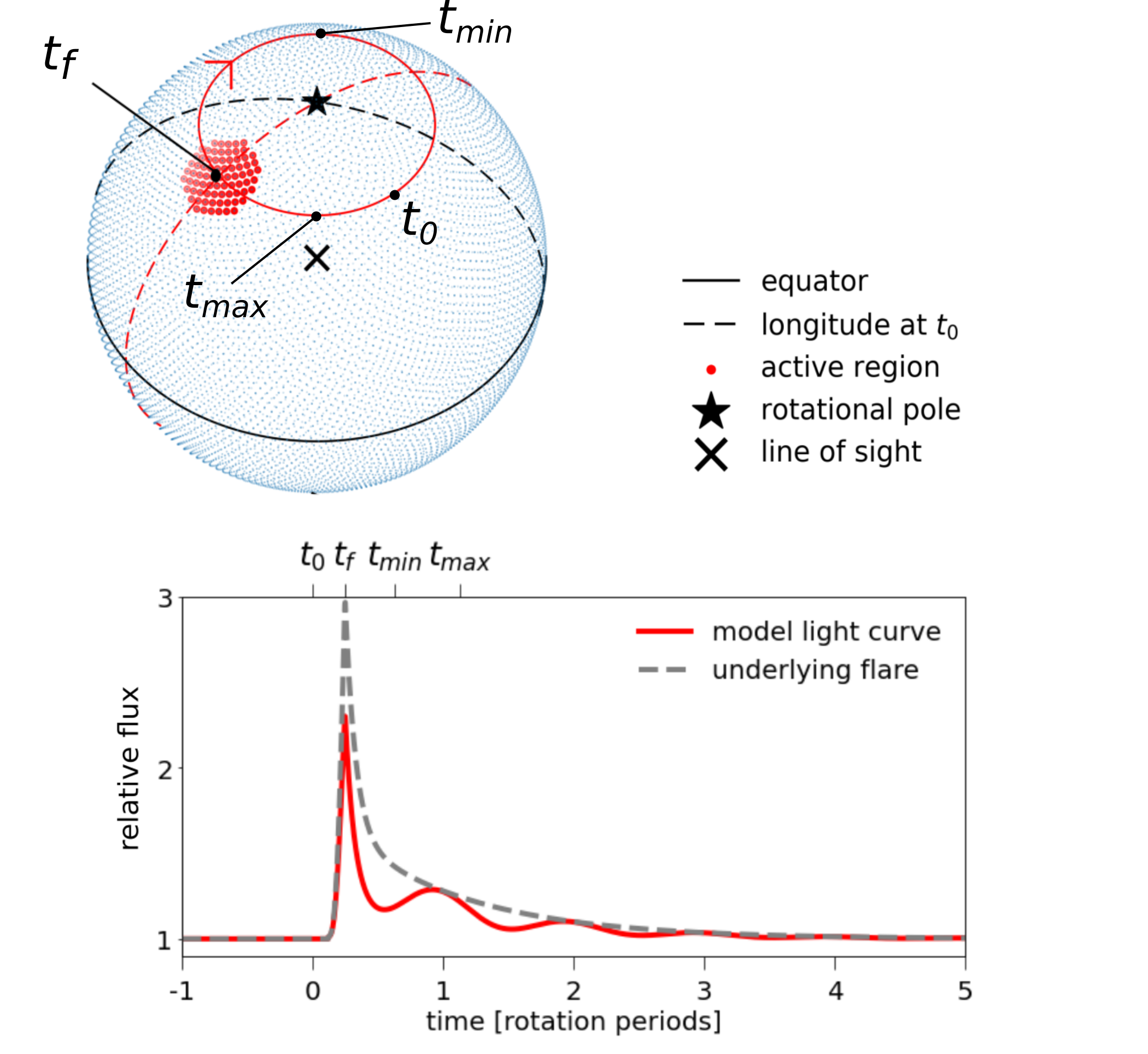}
    \caption{Direct flare localization from optical light curves. The star (blue sphere) rotates as the active region (red dots) erupts in a flare. Consequently, the flare light curve (grey dashed line in the bottom panel) is modulated geometrically in phase with rotation (red line). If the inclination of the stellar rotation axis is known, the location of the flaring region can be unambiguously recovered. We assume that the thermal optical emission captured by TESS originates from the dense lower strata of the stellar atmosphere, so that the flaring region can be seen as attached to the stellar surface.}
    \label{fig:ilin}
\end{figure}

With the launch of Kepler~\citep{borucki2010} and the Transiting Exoplanet Survey Satellite~\citep[TESS,][]{ricker2015}, systematic flare monitoring became feasible for hundreds of thousands of stars. In~\citet{ilin2021b}, we used TESS data from the first full sky scan of the mission to systematically search for rapidly rotating, fully convective stars, and found flares that could be located directly using the optical light curve of the star. 
We recovered four flares that lasted longer than a full rotation period on four fully convective M dwarfs.
These flares' light curves were modulated as the flare footpoints rotated in and out of view on the fast spinning stellar surface~(Fig.~\ref{fig:ilin}). We found that all \textit{flares occurred at latitudes closer to the rotational poles of the stars than to the equator}.
Except for an independent measurement of the inclination of the stellar rotation axis, required to lift the latitude-inclination degeneracy of the flaring region -- the light curve contained enough information to constrain the location of the flare on the stellar surface within a few degrees of uncertainty under the assumptions made about the flare temperature and the regions of optical emission.
As TESS, and soon PLATO~\citep{rauer2014}, will continue the high-cadence monitoring of flaring stars, this method can be applied to a steadily growing high quality data set.

\begin{figure}
    \centering
    \includegraphics[width=\hsize]{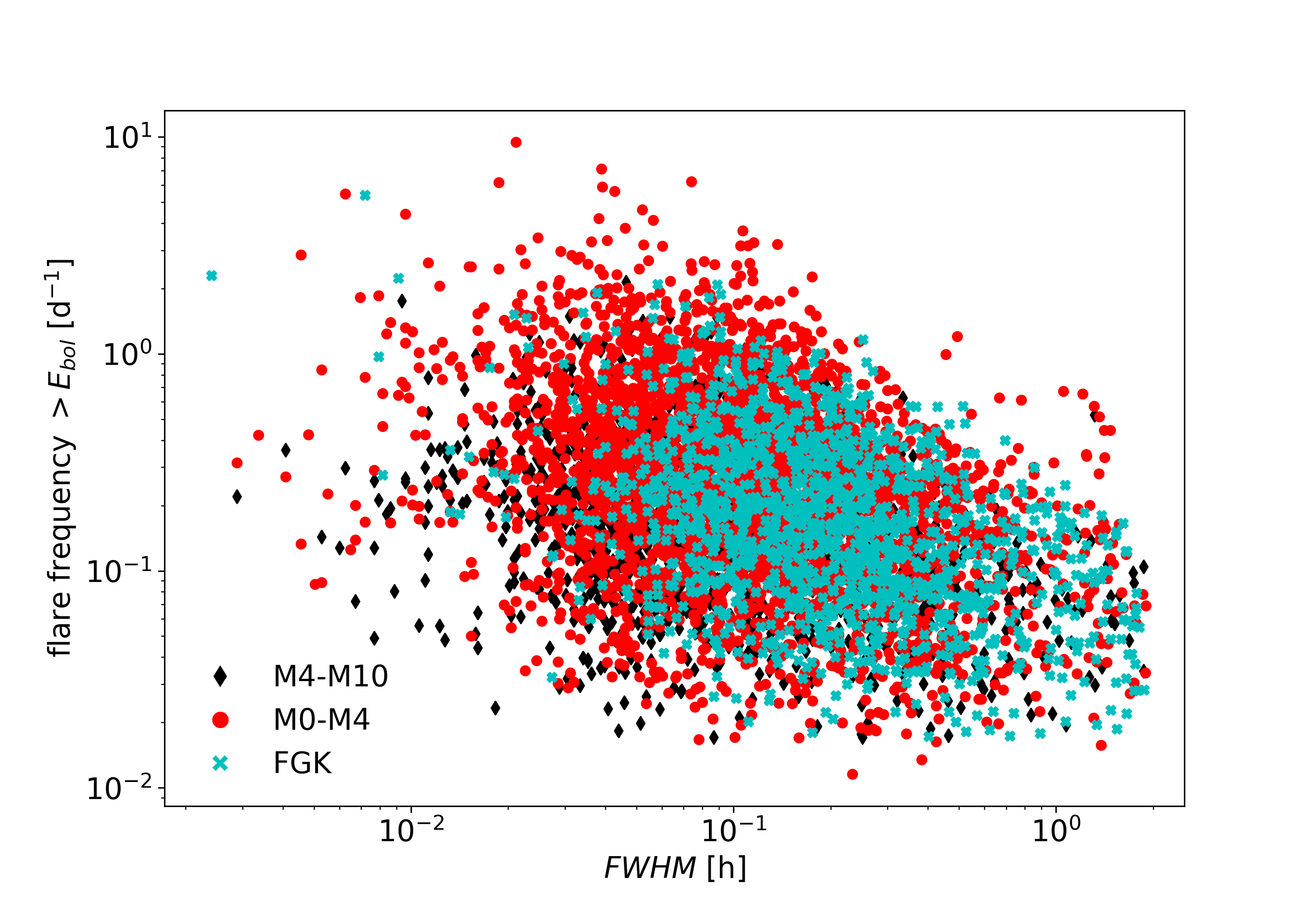}
    \caption{Flare frequency above bolometric flare energy $E_{bol}$ in flare stars with flares with measured $FWHM$. Note that the frequencies are for individual stars that were caught flaring within the first two Sectors of TESS Cycle 1 by~\citet{guenther2020}. Stars that flare but did not show any flares within the two first Sectors are not included.}
    \label{fig:guenther}
\end{figure}

\subsubsection{How many localizable flares will TESS observe?}
The method we presented in~\citet{ilin2021b} requires the star to rotate rapidly, and flare actively, so that an event than spans more than a full rotation period can possibly occur. 
Flares that last longer than 12 h are uncommon but not exceedingly rare.
Based on the TESS sample of fully convective dwarfs, we estimate that flares suited for localization analysis occur once every 120 years per star~\citep{ilin2021b}. This number will be lower for earlier type stars because they spin down faster~\citep{barnes2003,reiners2008}, losing their flaring activity at earlier ages~\citep{hilton2010,chang2015,ilin2019,ilin2021}. 
The flares in~\citet{ilin2021b} had TESS band energies around $10^{34}$ erg, and flare durations above half the maximum flare flux ($FWHM$) of about $1-1.5$ h. 

In Fig.~\ref{fig:guenther}we show how often flares above a certain energy occur as a function of that flare's $FWHM$. Gradual flares with $FWHM > 1$ h in actively flaring stars are fewer than impulsive flares, and these flares' energies tend to be high, which in turn lowers their frequencies among all flares (i.e., the upper right region in Fig.~\ref{fig:guenther} is mostly empty). Using the data from~\citet{guenther2020}, we can estimate how many localizable flares can be found in one full sky scan conducted with TESS.

Observations of young clusters show that stars are born with a wide distribution of stellar rotation rates~\citep[][and references therein]{johnstone2021}. 
At Pleiades age (125 Myr), about $30\%$ of stars with spectral types later than F5 and earlier than M6 rotate at $P_{rot}<12$~h, and $>75\%$ of them are fully convective~\citep{rebull2016}. 
At Praesepe/Hyades age (600-800 Myr), the mostly fully convective $P_{rot}<12$ h stars comprise only about $6\%$ of stars~\citep{douglas2019}.
Such fast rotators all reside in the saturated activity regime, and thus are expected to be actively flaring~\citep{doorsselaere2017, clarke2018}.

Assuming that the flaring stars in Fig.~\ref{fig:guenther} tend to be young, we can estimate that a similar fraction of these stars are rapidly rotating, and adopt $1\%$ as a lower limit. We can also take into account as a lower limit that about $1\%$ of stars are usually found flaring within one full-sky scan of TESS~\citep{tu2021, guenther2020}. From the flare data in~\citet{guenther2020}, we obtain a lower order of magnitude estimate of $0.01$ flares per day with sufficiently large $FHWM$, which yields an approximate lower limit of $10^{-6}$ localizable flares per day, or one flare every $\sim2 700$ years of observations per star. 
However, since TESS observes $200\,000$ mostly FGKM stars per full sky scan for at least $\sim 25$~d each, each two years of TESS operations would yield at least $\sim 5$ localizable flares, which is consistent with the four flares detected by \citet{ilin2021b} because the majority of the five are expected in the fully convective M dwarf regime based on the distribution of rotation rates.

\section{Implications of spatially non-uniform flare location distributions}
\label{sec:implications}
In the following sections, we discuss what opportunities and new questions locating flares on otherwise difficult to spatially resolve stellar surfaces opens up both for stellar and exoplanetary research.

\subsection{Localize flares and active regions to understand stellar magnetic fields}\label{sec:mf}

The connection between small scale surface and global magnetic fields of low mass stars is poorly understood. Flares, and in particular large stellar superflares, are unique probes of both: They occupy a volume that spans a wide range of atmospheric heights, and they are highly localized on the stellar sphere. Observing a flare at a known location on the star illuminates the dynamics of the strong small-scale surface magnetic field, and in the case of large superflares, ties it to high altitudes in the corona~\citep{benz2010}.

The triggering re-connection event is located in the corona, the subsequent energy release penetrates all layers of the stellar atmosphere through to the photospheric footpoints. Associated particle emission in the form of CMEs and energetic particle events (EP) will inevitably interact with the large scale field, which may confine, slow down, deflect or otherwise affect what eventually leaves the star along with the high energy radiation from the flare itself, which ultimately has an effect on the mass and angular momentum loss rates of the star~\citep{alvaradogomez2018,alvaradogomez2019,kay2019}. 

Prospectively, we can combine multi-wavelength time series observations that reveal how different atmospheric strata are affected by a flare~\citep{kowalski2013, maehara2021} with measurements of its latitude and longitude to reconstruct the full three-dimensional time evolution of flares on stars other than the Sun. 
\subsection{Localize flares to understand the impact of stellar activity on habitable zone planets}\label{sec:planets}

In Section~\ref{sec:systematic}, we have seen that young M dwarfs are favorable targets for the localization of large flares on the stellar surface. 
M dwarfs are of special interest in the context of habitability of exoplanets, because a. they are prime targets for transiting exoplanet surveys, b. their habitable zones are much closer to the star than in solar-type stars and c. they show high and persistent magnetic activity~\citep{shields2016}.
In these stars, large flares, and associated CMEs and EPs can alter and even evaporate the atmospheres of habitable zone planets.

Current models of atmospheric chemistry and evaporation do not take into account latitude distributions or assume solar ones~\citet{tilley2019, chen2021}, which is contrasted by observations that suggest that large flaring regions reside at much higher latitudes than on the Sun~\citep{wolter2008,ilin2021b}.
This is noteworthy because high latitude flares may have a different contribution to exoplanetary space weather than equatorial flares, especially CMEs and EPs.

Our analysis suggests that the optically thick blackbody emission that arises from the photospheric footpoints is attenuated by $40-80\%$ at the planet when flares are placed closer to the poles than to the equator~\citep[assuming the star-planet system is spin-orbit aligned,][]{ilin2021b}. However, it is the X-ray and UV emission, which is effective in altering the composition of or evaporating exoplanetary atmospheres, and affects biological life~\citep[see][and references therein]{airapetian2020}. It originates from a volume within the upper atmosphere of the star, and tends to be optically thin, so that the attenuation observed in the optical regime is not expected to occur.

On the other hand, solar EPs and CMEs are ejected in a cone radially outward from the locus of eruption~\citep{fisher1984,xie2004}. It is, however, an open question to what degree this holds true for other stars, and in particular stars with strong large scale fields that are also the stars that produce the most intense flares. 
Models have shown that such strong fields can confine, slow down, deflect and fragment the emitted magnetized plasma~\citep{alvaradogomez2018, alvaradogomez2019, kay2019}. If flares, however, occurred closer to the poles, where the prefentially dipolar component of the magnetic field has more open field lines, CMEs might escape more easily in the polar direction, and leave a planet in the equatorial plane largely unaffected. 
      
If flaring regions preferentially reside at certain latitudes on the stellar surface, spin-orbit alignment becomes an important ingredient in the assessment of stellar space weather. 
The above considerations assumed that the planet revolves around the rotational (and magnetic) equator of the star.
However, although we expect this to be true for planets that form in the protoplanetary disk and are left undisturbed in this position after formation, the research on spin-orbit alignment is still in its infancy, especially for terrestrial planets around low mass M dwarfs~\citep{mazeh2015,hirano2020,stefansson2020}.

\section{Summary}\label{sec:summary}
In the past, flares could only be located on the Sun directly, or by serendipitous discovery on a handful of stars. 
Methods available for stellar (magnetic) surface mapping like spectropolarimetry work best on large, stable structures like long lived spots or coronal loops, but usually mask more dynamic, smaller scale processes like flares. 
Asteroseismology and light curve inversion can also be used, both in isolation and combined, to infer the latitudes of starspot groups. However, the recovery of transient events like flares has yet to be demonstrated with these techniques.

With the rapid accumulation of long-term high cadence monitoring data of thousands of stars with Kepler, TESS, and soon PLATO, systematic flare localization becomes observationally more accessible, in particular for young, rapidly rotating, low mass stars.
The light curves of large, gradual flares obtained by these missions are modulated as the flaring region rotates in and out of view on the stellar surface. The morphology of the modulation can be used to directly infer the region's latitude and longitude.

Localizing flares constrains models of stellar magnetic fields, and helps us understand the interplay between surface magnetic fields and the overlying large-scale fields. When combined with multi-wavelength observations, the full spatio-temporal reconstruction of flares becomes a true possibility.

Finally, we have seen that we need to know the locus of flares and associated CMEs and EPs to accurately estimate the impact of these events on planets in close orbits around active low mass stars. As a consequence, stellar obliquity emerges as one of the key star-planet system characteristics.


\section*{Acknowledgments}
We thank the referee for their thorough reading and comments on the manuscript, and  Sebastian J. Pineda for helpful comments and discussion at he XMM-Workshop on CME escape.
EI acknowledges support from the German National Scholarship Foundation. KP acknowledges support from the German Leibniz Community under grant P67/2018. We made use of numpy~\citep{numpy2020} and pandas~\citep{pandas2010,pandas2020software}.

\subsection*{Author contributions}

All authors fulfil the criteria for authorship as stated in the AN Instructions for Authors. EI and KP were responsible for the design, the data analysis and interpretation in this work, and JDAG substantially contributed to the concept of the work, and the interpretation of the data.

\subsection*{Conflict of interest}
The authors declare no potential conflict of interests.

\bibliography{flareloci}
\end{document}